\documentclass[a4paper]{jpconf}
\usepackage{graphicx}

\begin{document}

\title{The influence of local majority opinions on the dynamics of the Sznajd model}

\author{Nuno Crokidakis}

\address{Departamento de F\'isica, PUC-Rio, Rio de Janeiro, Brazil}

\ead{nuno.crokidakis@fis.puc-rio.br}

\begin{abstract}
In this work we study a Sznajd-like opinion dynamics on a square lattice of linear size $L$. For this purpose, we consider that each agent has a convincing power $C$, that is a time-dependent quantity. Each high convincing power group of four agents sharing the same opinion may convince its neighbors to follow the group opinion, which induces an increase of the group's convincing power. In addition, we have considered that a group with a local majority opinion (3 up/1 down spins or 1 up/3 down spins) can persuade the agents neighboring the group with probability $p$, since the group's convincing power is high enough. The two mechanisms (convincing powers and probability $p$) lead to an increase of the competition among the opinions, which avoids dictatorship (full consensus, all spins parallel) for a wide range of model's parameters, and favors the occurrence of democratic states (partial order, the majority of spins pointing in one direction). We have found that the relaxation times of the model follow log-normal distributions, and that the average relaxation time $\tau$ grows with system size as $\tau \sim L^{5/2}$, independent of $p$. We also discuss the occurrence of the usual phase transition of the Sznajd model.

\end{abstract}

%#########################################################################

\section{Introduction}

In the last thirty years the dynamics of agreement/disagreement has been extensively studied by physicists, which is nowadays an active branch of \textit{Sociophysics} \cite{galam_book,castellano,stauffer_review}. Agreement is one of the most important features of social dynamics. Everyday life presents many situations in which it is necessary for a group to reach shared decisions. Agreement makes a position stronger, and amplifies its impact on society. Another interest in social phenomena is its theoretical motivation. In fact, several models present long-range correlations, power-law behavior and order-disorder phase transitions, among other typical features of physical systems \cite{galam_book,castellano,stauffer_review}.

Among the most studied opinion models in the last years we highlight the Galam models \cite{galam_review}, the voter model \cite{krapivsky_book} and the Sznajd model \cite{sznajd}. In special we are interested in the Sznajd model, that was initially defined on an one-dimensional ring, but was also considered on regular lattices in two and three dimensions, on complex networks and other topologies, and was applied to several practical problems like finance, marketing and politics (for a review, see \cite{sznajd_review}).

Despite the above applications of the Sznajd model, it does not consider important social features. The dynamics of social relationships in the real world shows a large number of details that are commonly neglected in some theoretical models. In order to introduce more realistic features, we have considered in this work a convincing-power mechanism that limits the capacity of persuasion of the agents in the Sznajd model, as discussed in \cite{sznajd_meu}. In addition, we have considered that not only groups of four agents sharing the same opinion can persuade their neighbors, but also groups with a local majority opinion can be persuasive in the opinion formation process. It is expected that the inclusion of such ingredients in the Sznajd model turns it closer to a real social system, where not only the number of individuals with same opinion matters, but also their persuasiveness. Furthermore, it has been discussed that a local majority opinion can be considerably influent on the dynamics of public debates \cite{galam_review}.

This work is organized as follows. In Section 2 we present the model and define their microscopic rules. The numerical results as well as the finite-size scaling analysis are discussed in Section 3. Finally, our conclusions are presented in Section 4.

%#########################################################################

\section{Model}

The model considered in this paper is an extension of a recente work \cite{sznajd_meu}. We have considered the Sznajd model defined on a square lattice with linear size $L$ and periodic boundary conditions \cite{adriano}. Each agent on a given lattice site $i$ carries an Ising-like variable $s_{i}=\pm 1$, representing the two possible opinions (yes or no, for example) in a given subject. The initial configuration of opinions is given by a fraction or density $d$ ($1-d$) of $s=+1$ ($s=-1$) opinions. Furthermore, an integer number ($C$) labels each player and represents its \textit{convincing power} across the community. The convincing power is introduced as a score which is time dependent. The agents start with a random distribution of the $C$ values, and during the time evolution the convincing power of each agent can change according to its capacity of persuasion, following the model's rules. One time step in our model is defined by the following rules:

\begin{enumerate}
\item We randomly choose a 2 $\times$ 2 plaquette or group of four neighbors.

\item If all four center spins are parallel, we calculate the average convincing power $\bar{C}$ of the plaquette:
\begin{eqnarray} \nonumber
\bar{C} = \frac{1}{4}\sum_{i=1}^{4}C_{i}~,
\end{eqnarray}
where each term $C_{i}$ represents the convincing power of one of the plaquettes' agents.
\item Then, we compare the average convincing power $\bar{C}$ with the convincing power of each one of the 8 sites neighboring the plaquette. Thus, given a plaquette neighbor $j$ with convincing power $C_{j}$, this individual $j$ will follow the plaquette opinion if $\bar{C}>C_{j}$. In this case, the convincing power of each agent in the 2 $\times$ 2 group is increased by 1.
\item If not all four spins are parallel, we verify if there is a majority opinion in the group (3 up/1 down spins or 1 up/3 down spins). In this case, we apply the above rule with probability $p$, i.e., we calculate the average convincing power $\bar{C}_{{\rm maj}}$ of the agents with the majority opinion, and compare $\bar{C}_{{\rm maj}}$ with the convincing power of each one of the 8 sites neighboring the plaquette. For each persuasion, the convincing power of each agent with majority opinion in the 2 $\times$ 2 group is increased by 1.

\end{enumerate}
 
Thus, the convincing powers of the agents neighboring a given group act in the dynamics as a conviction, in a way that if a given plaquette neighbor $j$ has a convincing power $C_{j}>\bar{C}$ he/she will not be persuaded to change opinion. In this case, the above-mentioned rules define a competition among a group of agents sharing the same opinion and the conviction of the agents neighboring the group. In addition, we will consider also the influence of the majority opinion in the group. In this case, with probability $p$ an opinion which is the majority one in a given group may be adopted by the group's neighbors, since this local majority opinion is shared by a high convincing-power group.

The objective of the agents in this dynamics is to convince their neighbors of their opinion. One can expect that, if a certain group of agents convinces many other agents, the persuasion ability of each agent in the group increases. Thus, the inclusion of the convicing powers in the model may capture this real-world characteristic. In addition, as considered in Galam's models \cite{galam_review}, a local majority opinion in a discussion group considerably affects the dynamics of a public debate. For the limiting case $p=0.0$, we recover the model studied in \cite{sznajd_meu}.

%#########################################################################

\section{Numerical Results}

In the simulations, the initial values of the agents' convincing powers follow a gaussian distribution centered at $0$ with standard deviation $\sigma=5$ \footnote{As discussed in \cite{sznajd_meu}, the variation of $\sigma$ does not affects quantitatively the results of the model, which also occurs in the present formulation.}. We can start studying the time evolution of the quantity analogous to the ``magnetization per spin", i.e., 
\begin{equation}
m(t) = \frac{1}{N}\sum_{i=1}^{N}s_{i} ~,
\end{equation}
%%%%%%%%%%%%%%%%%%%%%%%%%%%%%%%%%%%%%%%%%%%%%%%%%%%%%%%%%%%%%%%%%%%%%%%%%%
\begin{figure}[t]
\begin{center}
\vspace{0.2cm}
\includegraphics[width=17pc]{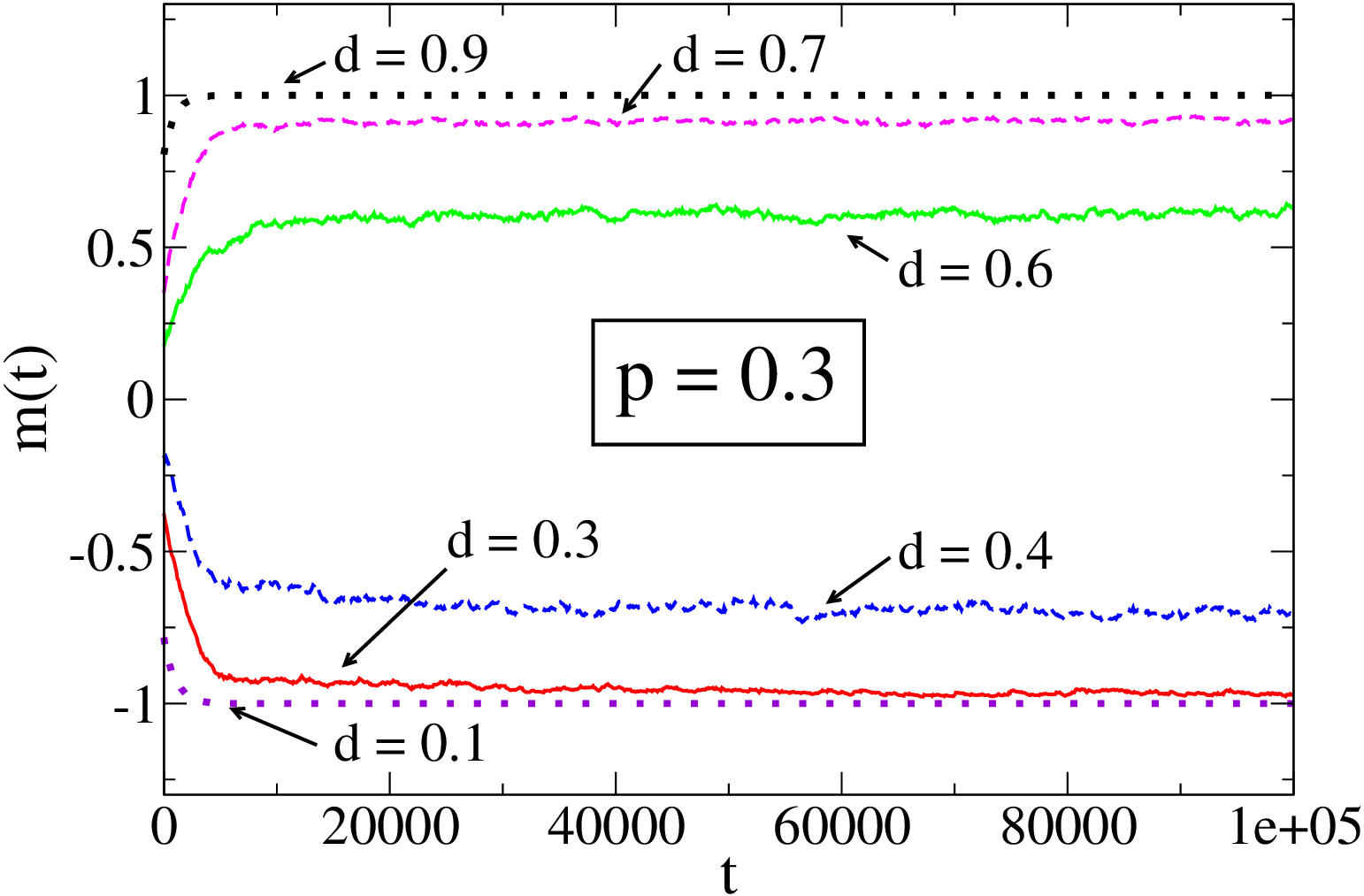}
\hspace{0.2cm}
\includegraphics[width=17pc]{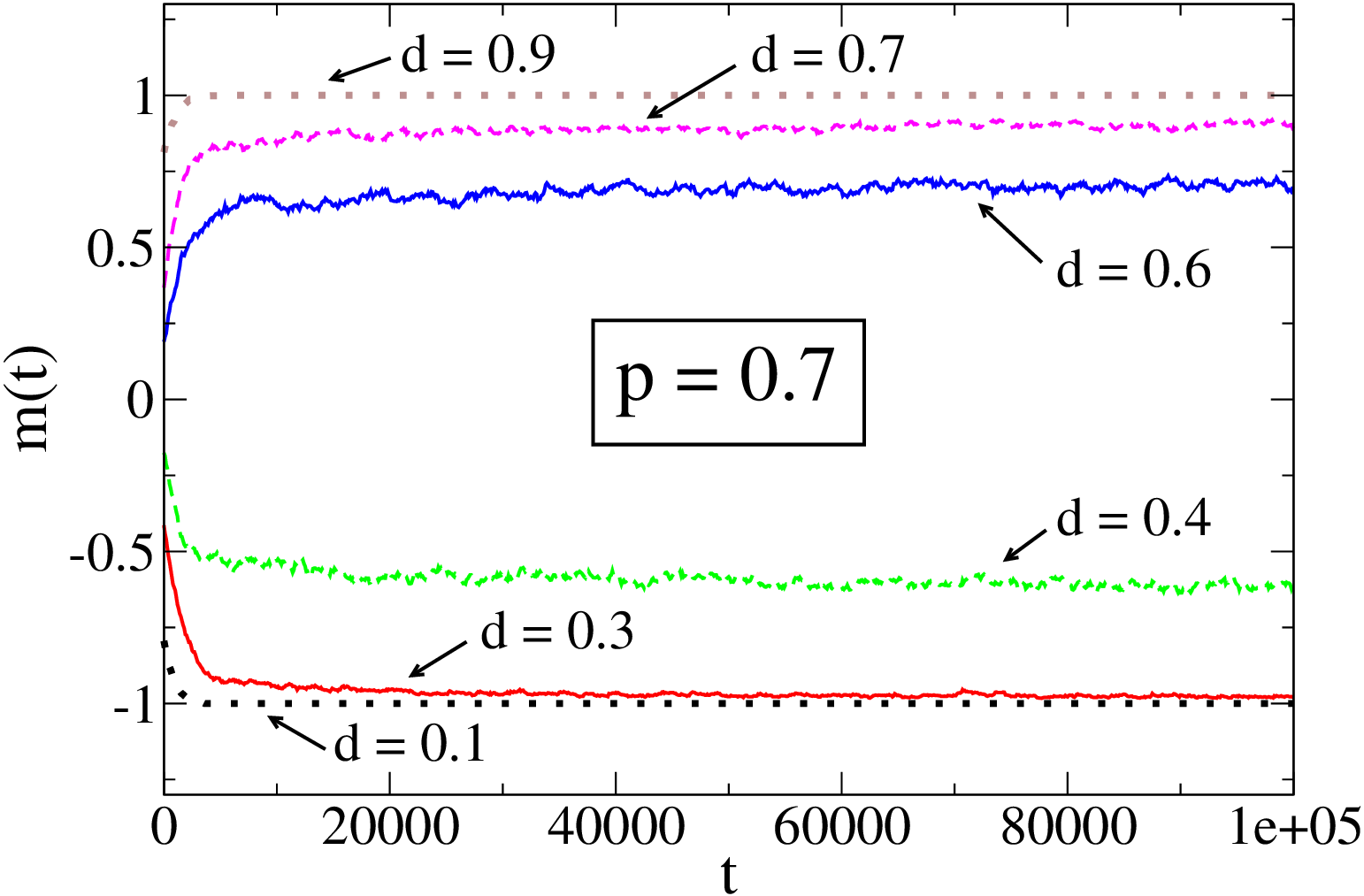}
\end{center}
\caption{Time evolution of the magnetization per spin for $L=53$, some values of the initial density $d$ of $s=+1$ opinions and two different values of the probability $p$, namely $p=0.3$ (left side) and $p=0.7$ (right side). We can observe steady states with $|m|=1$ (consensus) as well as with $|m|<1$ (no consensus). Each curve is a single realization of the dynamics.}
\label{fig1}
\end{figure}
%%%%%%%%%%%%%%%%%%%%%%%%%%%%%%%%%%%%%%%%%%%%%%%%%%%%%%%%%%%%%%%%%%%%%%%%%%

\noindent
where $s_{i}=\pm 1$ and $N=L^{2}$ is the total number of agents on the square lattice. We exhibit in Figure \ref{fig1} the magnetization per spin $m$ as a function of the number $t$ of time steps for $p=0.3$ (left side) and $p=0.7$ (right side) and some values of the initial density $d$ of $s=+1$ opinions. Observe that the time needed for the system to reach the steady states does not vary considerably from one case to another. One can also see in Figure \ref{fig1} that situations with $|m|<1$ in the long-time limit (steady state) emerge spontaneously from the dynamics, as in the $p=0.0$ case \cite{sznajd_meu}. This situation does not occurs in the original two-dimensional Sznajd model, where only absorbing states with $|m|=1$ (consensus states) are obtained in the steady states \cite{adriano}. In our case, such full-consensus states with all spins pointing up or down are possible only for large and small values of $d$, respectively, as we can also see in Figure \ref{fig1}. Thus, the consideration of the influence of a local majority opinion does not favor consensus states, usually related to dictatorships \cite{adriano}, making stable steady states with $|m|<1$, that can be related to a democracy. This makes the Sznajd model more realistic. One can also see in Figure \ref{fig1} that there are transitions between a phase where the full-consensus states with $|m|=1$ are always reached and another phase where this consensus never occurs. This fact is more clear in Figure \ref{fig2}, where we exhibit the time evolution $m(t)$ for a large initial density of $+1$ opinions ($d=0.9$) and for two distinct values of $p$, namely $p=0.5$ and $p=0.9$. In this case, the increase of the influence of groups with a majority opinion (i.e., the increase of the probability $p$) leads the system to steady states with $|m|<1$. These results suggest that the usual phase transition of the Sznajd model may not occurs in our model for large values of $p$. We will analyze this phase transition in more detail in the following.

%%%%%%%%%%%%%%%%%%%%%%%%%%%%%%%%%%%%%%%%%%%%%%%%%%%%%%%%%%%%%%%%%%%%%%%%%%
\begin{figure}[t]
\begin{center}
\vspace{0.5cm}
\includegraphics[width=18pc]{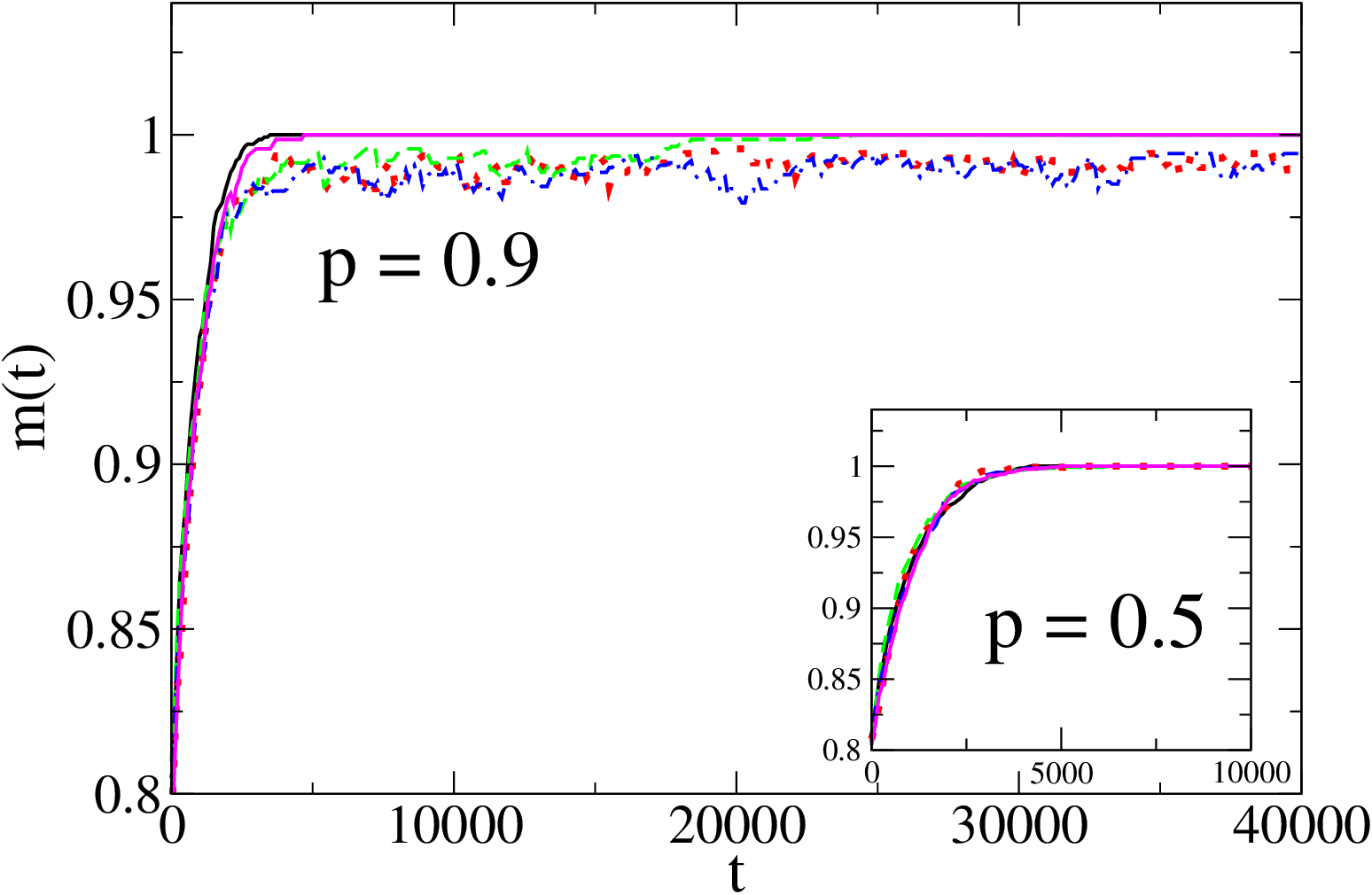}
\end{center}
\caption{Time evolution of the magnetization per spin for $L=53$, $d=0.9$ and two distinct values of $p$, namely $p=0.9$ (main plot) and $p=0.5$ (inset). We have considered 5 independent simulations for each value of $p$. One can see that even for a large initial density of $+1$ opinions, the increase of the influence of groups with a majority opinion does not necessarily lead the system to the absorbing state with $m=1$.}
\label{fig2}
\end{figure}
%%%%%%%%%%%%%%%%%%%%%%%%%%%%%%%%%%%%%%%%%%%%%%%%%%%%%%%%%%%%%%%%%%%%%%%%%%

%%%%%%%%%%%%%%%%%%%%%%%%%%%%%%%%%%%%%%%%%%%%%%%%%%%%%%%%%%%%%%%%%%%%%%%%%%
\begin{figure}[t]
\begin{center}
\vspace{0.5cm}
\includegraphics[width=18pc]{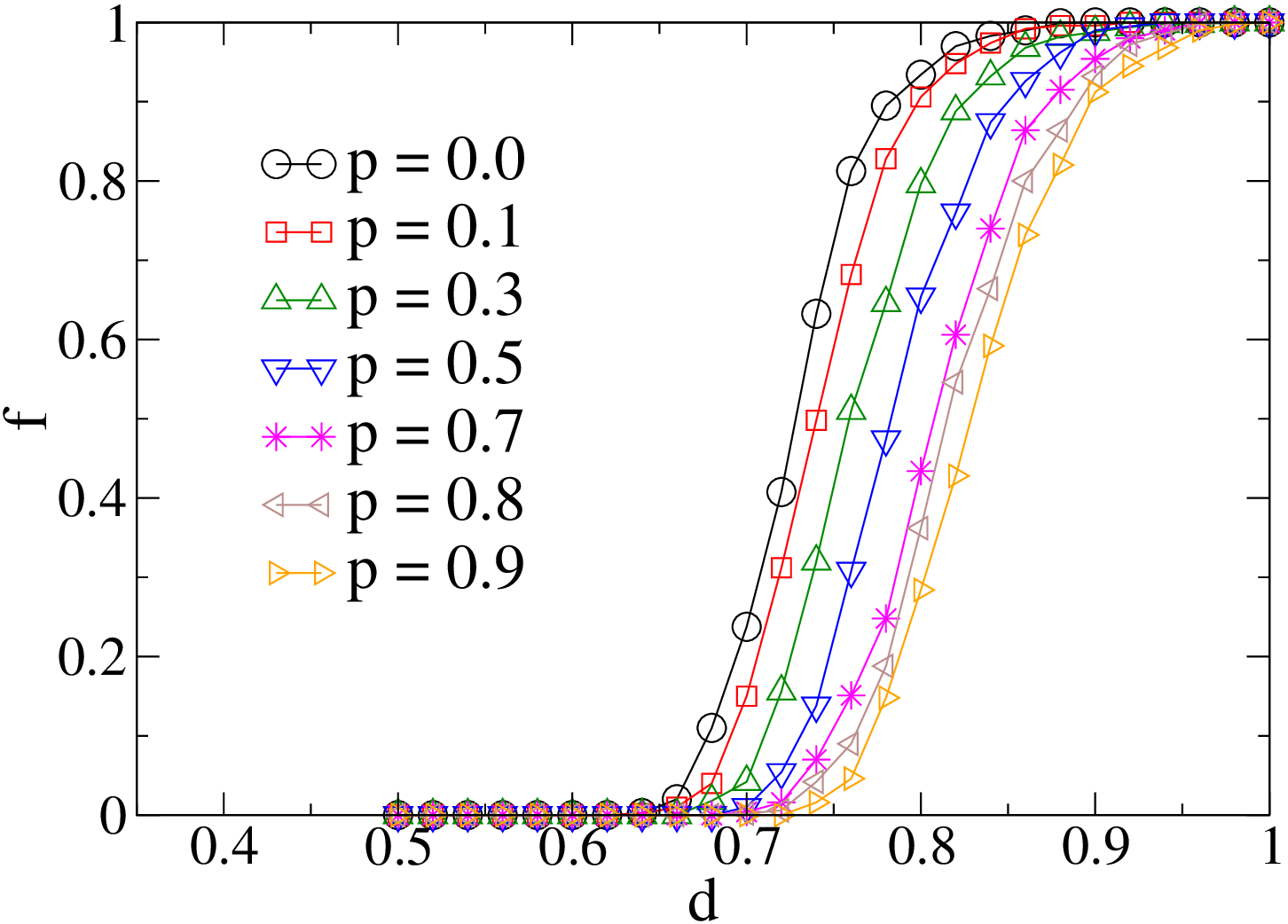}
\end{center}
\caption{Fraction $f$ of samples which show all opinions $+1$ when the initial density $d$ of $+1$ opinions is varied in the range $0.5\leq d\leq 1.0$, for $L=73$ and typical values of the probability $p$. One can observe the usual phase transition from $f=0$ to $f=1$ at values $d_{c}(p)$, at least for $p< 0.8$.}
\label{fig3}
\end{figure}
%%%%%%%%%%%%%%%%%%%%%%%%%%%%%%%%%%%%%%%%%%%%%%%%%%%%%%%%%%%%%%%%%%%%%%%%%%

\begin{figure}[t]
\begin{center}
\vspace{0.2cm}
\includegraphics[width=17.2pc]{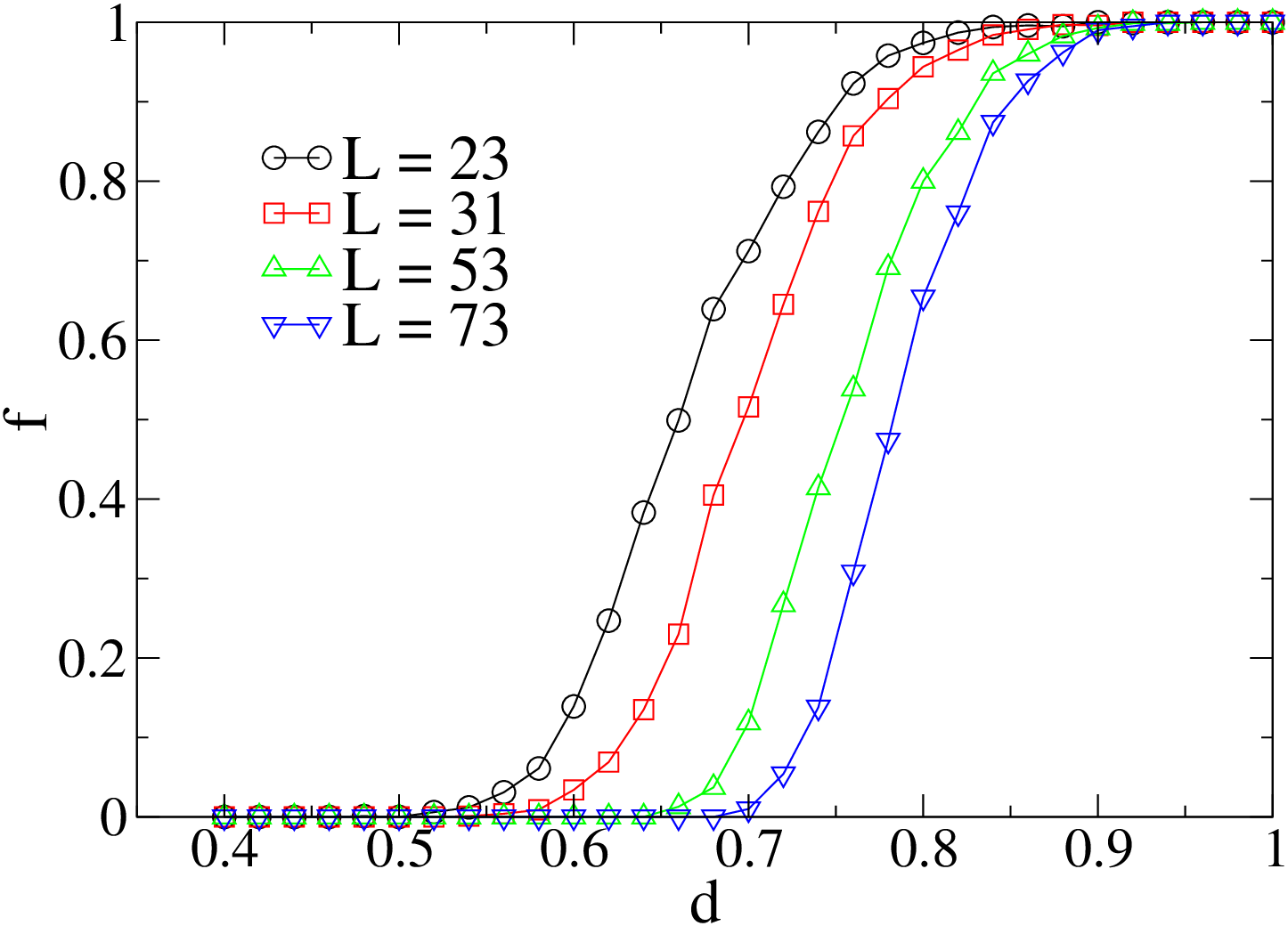}
\hspace{0.2cm}
\includegraphics[width=17pc]{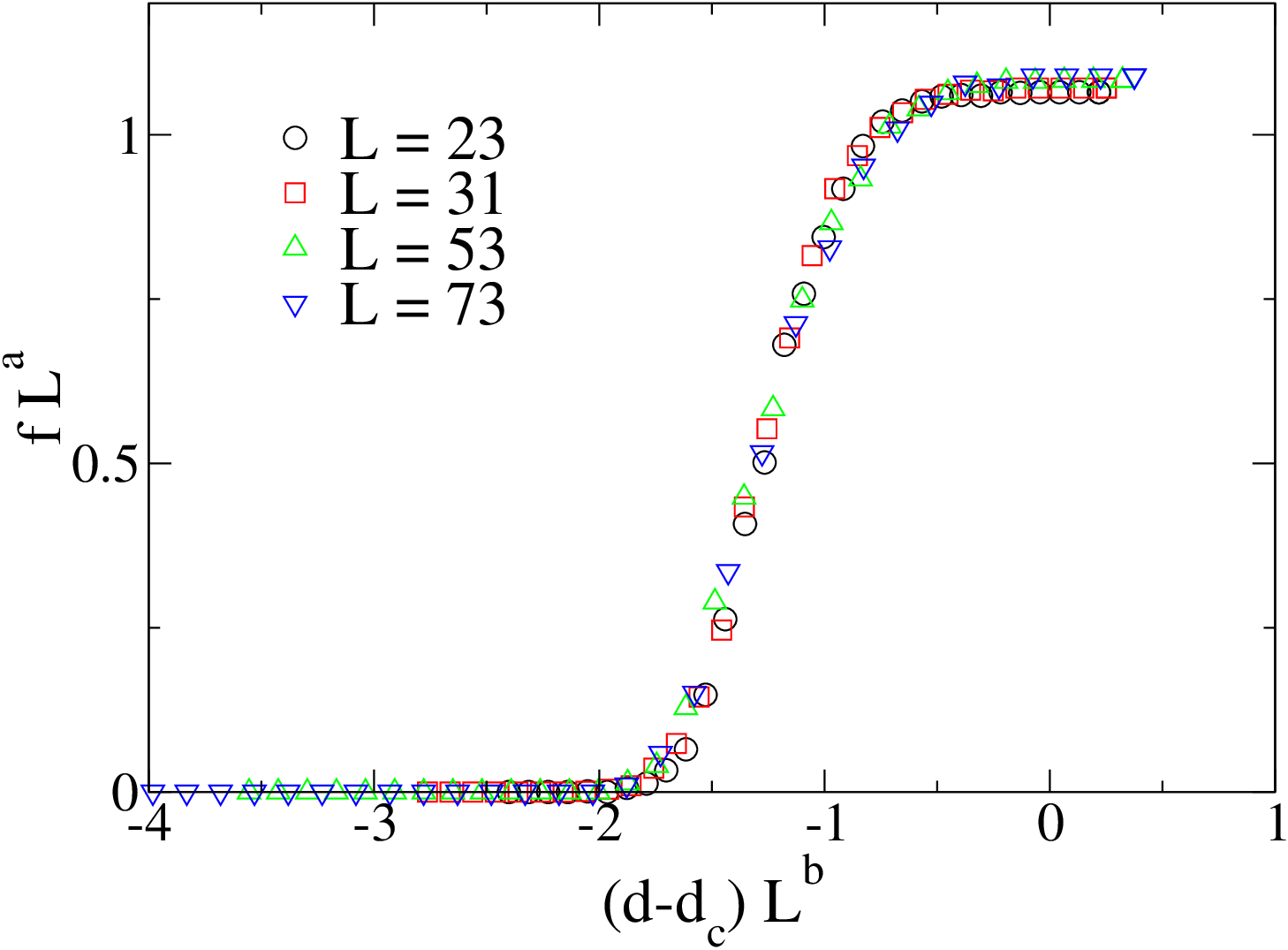}
\end{center}
\caption{Fraction $f$ of samples which show all opinions $+1$ when the initial density $d$ of $+1$ opinions is varied in the range $0.4\leq d\leq 1.0$, for $p=0.5$ and some values of the lattice size $L$ (left side). It is also exhibited the corresponding scaling plot of $f$ (right side). The best collapse of data was obtained for $a=0.02$, $b=0.47$ and $d_{c}=0.95$.}
\label{fig4}
\end{figure}

In order to better study the phase transition, we simulated the system for different probabilities $p$, and we measured the fraction $f$ of samples which show all opinions $+1$ when the initial density $d$ of $+1$ opinions is varied in the range $0.5\leq d\leq 1.0$. In other words, this quantity $f$ gives us the probability that the population reaches consensus, for a given value of the pair ($d,p$). The numerical results for $f$ as a function of $d$ for $L=73$ and typical values of $p$ are shown in Figure \ref{fig3}. Observe that for values of $p$ near $0.8$ the system reaches the value $f=1.0$ only for $d\to 1.0$, which suggests that there is no phase transition for large values of the probability $p$. For the formal characterization of the transition, we have to simulate the system for different lattice sizes $L$. We have considered $1000$ samples for $L=23$ and $L=31$, $700$ samples for $53$ and $500$ samples for $L=73$. An example is exhibited in Figure \ref{fig4} (left side), for $p=0.5$. In order to locate the critical point, we performed a finite-size scaling (FSS) analysis, based on the standard FSS equations \cite{sznajd_meu},
%%%%%%%%%%%%%%%%%%%%%%%%%%%%%%%%%%%%%%%%%%%%%%%%%%%%%%%%%%%%%%%%%%%%%%%%%%
%%%%%%%%%%%%%%%%%%%%%%%%%%%%%%%%%%%%%%%%%%%%%%%%%%%%%%%%%%%%%%%%%%%%%%%%%%

\begin{eqnarray}
f(d,L) & = & L^{-a}\;\tilde{f}((d-d_{c})\;L^{-b}) ~, \\
d_{c}(L) & = & d_{c}+a\;L^{-b} ~.
\end{eqnarray}
\noindent
 The result is shown in Figure \ref{fig4} (right side), and we have found that 
\begin{equation}
d_{c}(p=0.5)=0.95 \pm 0.01 ~,
\end{equation}
\noindent
in the limit of large L, with exponents $a\approx 0.02$, $b\approx 0.47$. Notice that the values of the critical exponents are similar to the ones observed in the $p=0.0$ case \cite{sznajd_meu}. Nonetheless, the value of $d_{c}$ is greater than the critical density for $p=0.0$, for which the numerical result is $d_{c}(p=0.0)=0.88 \pm 0.01$ \cite{sznajd_meu}. This result indicates that an important effect of the influence of local majority opinions is to shift the critical point for higher values, but the universality class of the model remains the same for different values of the probability $p$.

%%%%%%%%%%%%%%%%%%%%%%%%%%%%%%%%%%%%%%%%%%%%%%%%%%%%%%%%%%%%%%%%%%%%%%%%%%
\begin{figure}[t]
\begin{center}
\vspace{0.5cm}
\includegraphics[width=18pc]{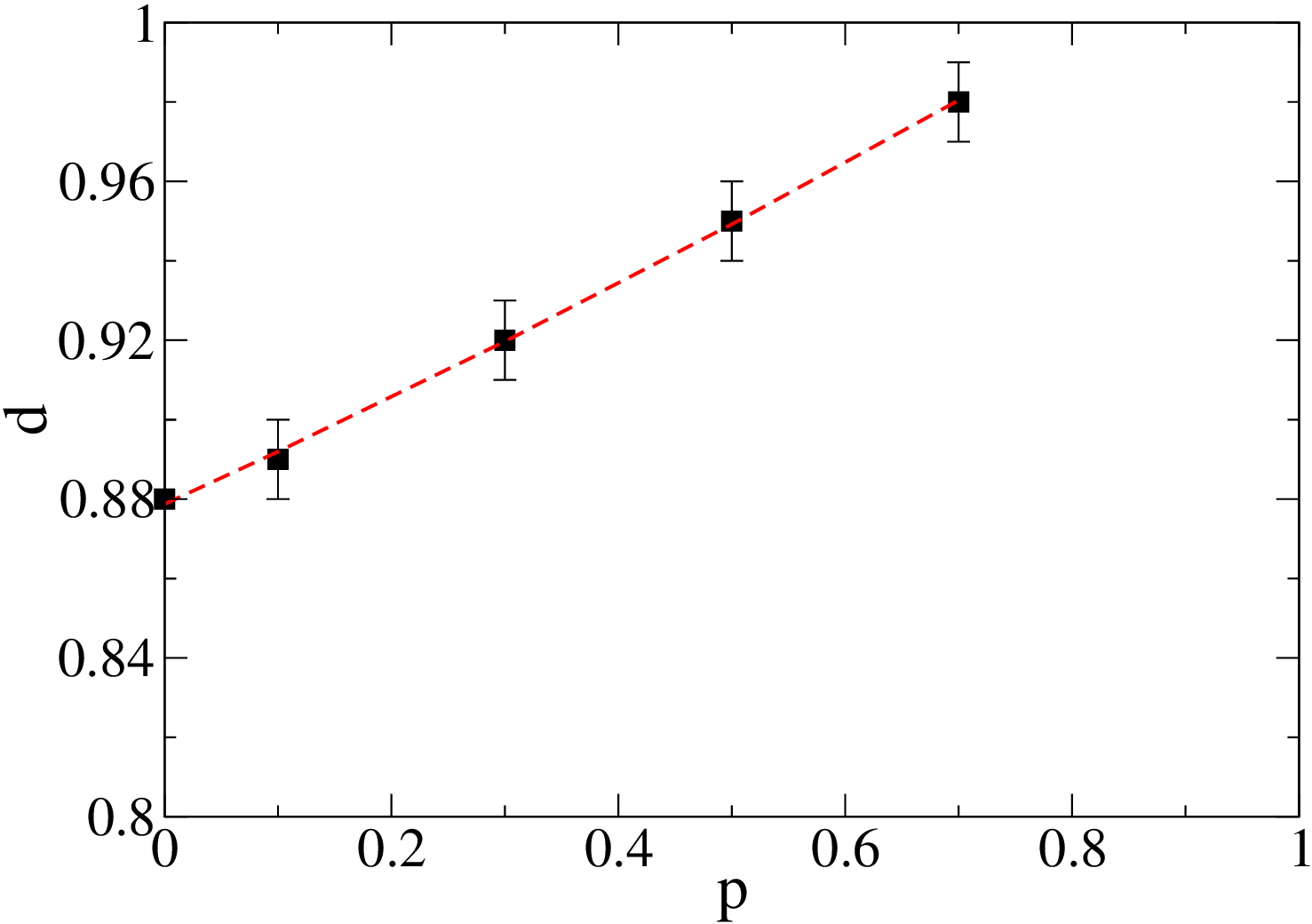}
\end{center}
\caption{Phase diagram of the model in the plane $d$ versus $p$. The points are numerical estimates of the critical densities $d_{c}(p)$, whereas the dashed line is just a guide to the eyes. In the region above the curve the system always reaches consensus with all opinions $+1$, whereas in the region below the curve this kind of consensus never occurs.}
\label{fig5}
\end{figure}
%%%%%%%%%%%%%%%%%%%%%%%%%%%%%%%%%%%%%%%%%%%%%%%%%%%%%%%%%%%%%%%%%%%%%%%%%%

In order to estimate the critical points $d_{c}(p)$ for other values of $p$, we repeated the above-discussed FSS analysis. The results suggest that we have $d_{c}(p)<1.0$ (within error bars) for $p\leq 0.7$, which indicates the occurrence of a phase transition, with similar exponents $a\approx 0.02$, $b\approx 0.47$, which confirms the universality class for distinct values of $p$. On the other hand, the best collapse of data for $p>0.7$ was obtained for $d_{c}=1.0$. In this case, we can not affirm if the phase transition occurs. Taking into account the critical densities $d_{c}(p)$ calculated for typical values of $p$, we exhibit in Figure \ref{fig5} the phase diagram of the model in the plane $d$ versus $p$. The points are the critical densities $d_{c}(p)$ estimated from the FSS analysis of the numerical results, whereas the dashed line is just a guide to the eyes. Considering the thermodynamic limit $L\to\infty$, the critical points $d_{c}$ separate a region where the full-consensus states with all opinions $+1$ (or all spins up) always occur, for each realization of the dynamics (above the frontier), from a region where this kind of consensus never occurs (below the curve).

%%%%%%%%%%%%%%%%%%%%%%%%%%%%%%%%%%%%%%%%%%%%%%%%%%%%%%%%%%%%%%%%%%%%%%%%%%
\begin{figure}[t]
\begin{center}
\vspace{0.2cm}
\includegraphics[width=16pc]{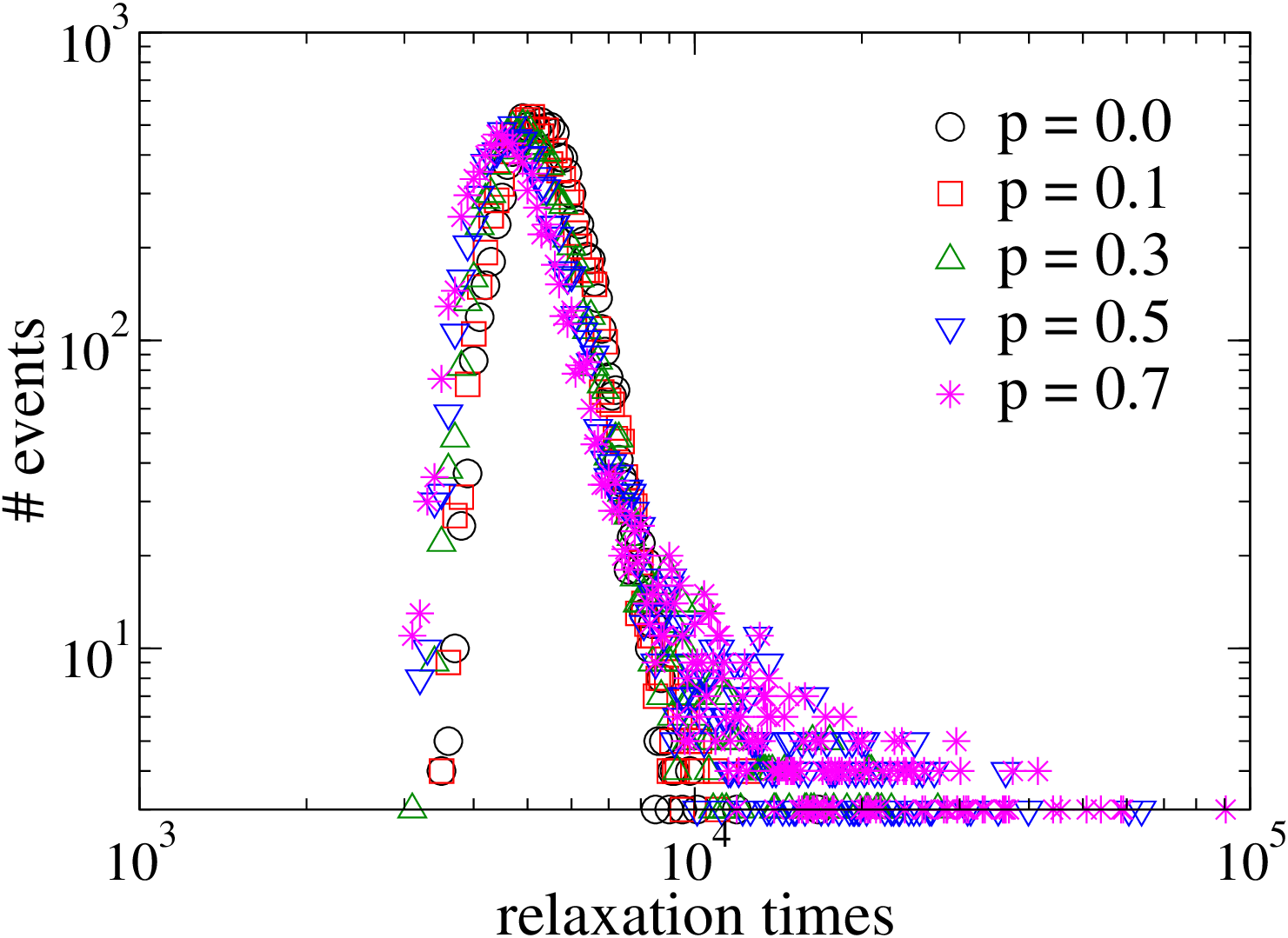}
\hspace{0.2cm}
\includegraphics[width=16pc]{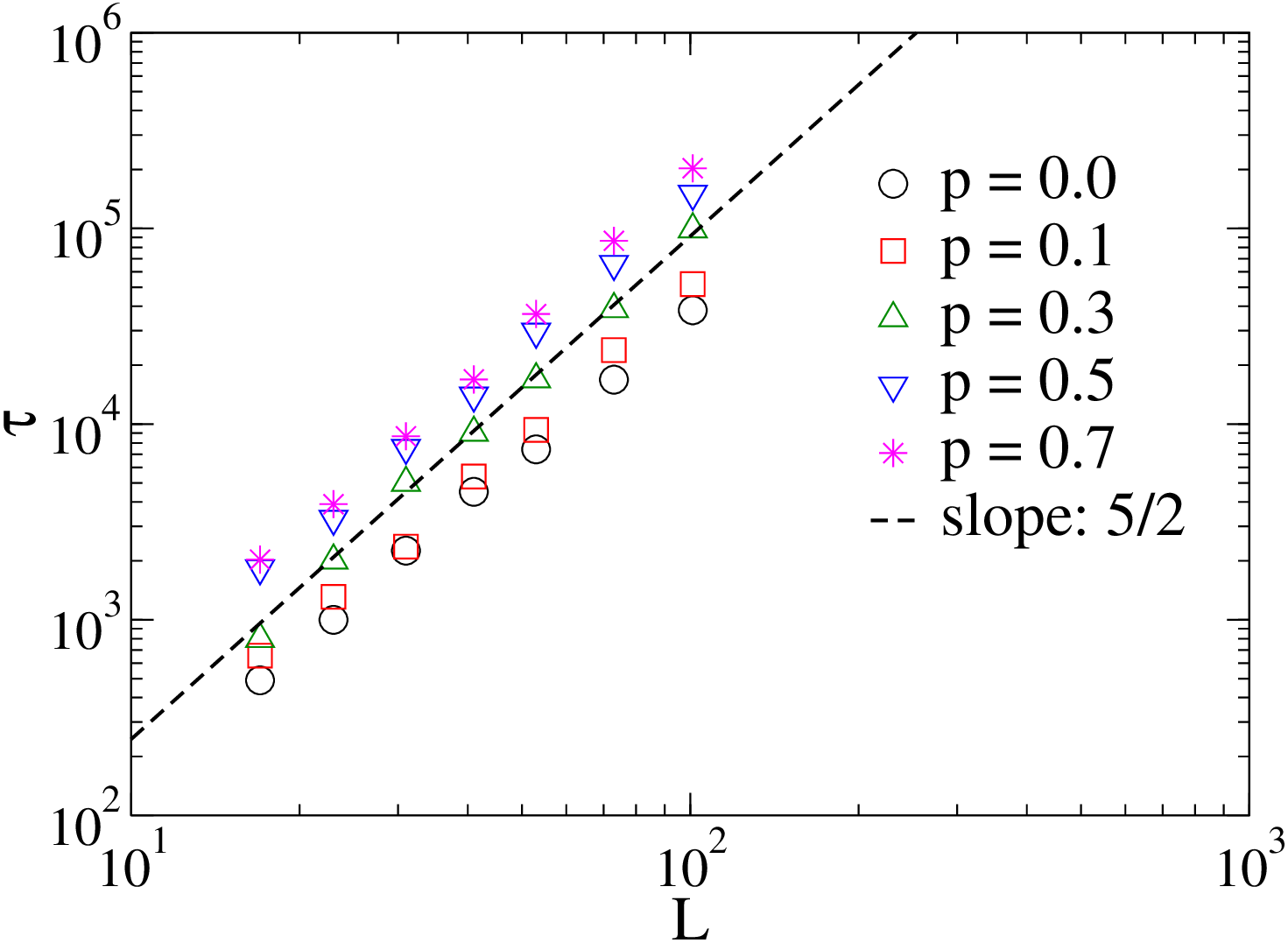}
\end{center}
\caption{Histogram of the relaxation times for $L=53$ and $d=0.9$ (left side). It is also exhibited the log-log plot of the average relaxation time $\tau$ as a function of the lattice size $L$ (right side). The dashed line has slope 5/2.}
\label{fig6}
\end{figure}
%%%%%%%%%%%%%%%%%%%%%%%%%%%%%%%%%%%%%%%%%%%%%%%%%%%%%%%%%%%%%%%%%%%%%%%%%%

We have also studied the relaxation times of the model, i.e., the time needed to find all the agents at the end having the $+1$ opinion. The distribution of the number of sweeps through the lattice, averaged over $10^{4}$ samples, needed to reach the fixed point is shown in Figure \ref{fig6} (left side), for typical values of the probability $p$. We can see that the relaxation time behavior is compatible with a log-normal distribution for all values of the probability $p$, which corresponds to the parabolas in the log-log plot of Figure \ref{fig6} (left side). In addition, the numerical results suggest that the relaxation times are not substantially affected by the probability $p$ in the central region (near the peaks), which is in agreement with the above qualitative discussion based on the results shown in Figure \ref{fig1}. The difference appears in the tail of the distribution, that presents larger relaxations times for increasing values of $p$. These tails are responsable for the difference among the average relaxations times $\tau$ for a given value of the lattice size $L$, as one can see in Figure \ref{fig6} (right side). The data are plotted in a log-log scale, and we can verify a power-law relation between those quantities in the form 
\begin{equation}
\tau \sim L^{5/2} ~.
\end{equation}
Notice that the above exponent is the same for all values of $p$, including the special case $p=0.0$ \cite{sznajd_meu}, and thus it is not affected by the influence of local majority opinions.

%#########################################################################

\section{Final Remarks}

In this work, we have analyzed the consequences of the introduction of two distinct mechanisms in the Sznajd model. The first one is the convincing power, that is introduced as a score ($C$) which is time dependent. The agents start with a gaussian distribution of the $C$ values, and during the time evolution, the convincing power of each agent changes according to its capacity of persuasion, following the model's rules. The second mechanism is the influence of local majority opinions. In this last case, if there is a local majority opinion in a group of 4 agents (3 up/1 down spins or 1 up/3 down spins), with probability $p$ these agents sharing the majority opinion can persuade their neighbors to adopt the mentioned opinion. We analyzed the influence of these two ingredients on the dynamics of the Sznajd model and on the opinion spreading.

A first consequence of the presence of those ingredients is the absence of consensus (all opinions $+1$ or all spins up) for a wide range of parameters. These states are obtained only for very small or very large initial densities $d$ of $+1$ opinions. In this sense, this formulation of the Sznajd model appears to be more realistic, since consensus states are usually related to dictatorships. On the other hand, states with magnetization per spin $|m|<1$ are desirable since they represent democracy-like situations \cite{sznajd_meu,adriano}. A second effect of the influence of the local majorities is that the critical density of up spins $d_{c}(p)$ shifts for higher values with increasing values of $p$. In addition, the usual phase transition of the Sznajd model occurs for all values of $p\leq 0.7$, but for $p>0.7$ we can not affirm if that phase transition occurs. In the cases where the transition occurs, the critical exponents are universal and they are independent of $p$.

As in the standard 2D Sznajd model \cite{adriano}, we found a log-normal distribution of the relaxation times to consensus. In addition, the average relaxation time $\tau$ depends on the lattice size in a power-law form, $\tau\sim L^{5/2}$, which is also independent of $p$.

%#########################################################################
\medskip

\section*{Acknowledgments}

The author acknowledges financial support from the Brazilian funding agencies FAPERJ and CAPES.

\smallskip

\end{document}